\documentclass{article}
\usepackage{amsmath}
\usepackage{amssymb,dsfont}

\def\pois#1#2{\left\{ {#1},{#2} \right\}}
\def\co{\Delta}
\def\alg{{\mathfrak g}}
\def\endpf{\begin{flushright}$\square$\end{flushright}}
\def\al{\alpha}
\def\be{\beta}
\def\ga{\gamma} 
\def\de{\delta}
\def\ep{\epsilon}
\def\back{\!\!\!\!\!\!\!\!\!\!\!\!\!\!\!\!\!}
\def\part#1#2{\frac{\partial {#1}}{\partial {#2}}} 
\def\la{\lambda}

\newtheorem{theorem}{Theorem}

\begin{document}

\title{Loop Coproducts}

\author{F. Musso\footnote{Departamento de F\'isica, Universidad de Burgos, 
E-09001 Burgos, Spain, email: fmusso@ubu.es}}

\date{}

\maketitle

\begin{abstract}
\noindent In this paper we show that if $A$ is a Poisson algebra equipped with a set of maps $\Delta^{(i)}_\la:A \rightarrow A^{\otimes \, N}$ 
satisfying suitable conditions, then the images of the Casimir functions of $A$ under the maps $\Delta^{(i)}_\la$ (that we call ``loop 
coproducts'') 
are in involution. Rational, trigonometric and elliptic Gaudin models can be recovered as particular cases of this result, and we show
that the same happens for the integrable (or partially integrable) models that can be obtained through the so called coproduct method.

On the other hand, this loop coproduct approach is potentially much more general, and could allow the generalization of the Gaudin algebras
from the Lie-Poisson to the Poisson algebras context and, hopefully, the definition of new integrable models.
\end{abstract}
 
\section{Introduction}

The Gaudin model was defined in the '70 by Gaudin \cite{Gaudin1,Gaudin2} as an integrable quantum spin chain model. The model was later 
cast in the framework of linear r-matrix formulation by Sklyanin \cite{Sk89} and Jur\v{c}o \cite{Jurco}. The linear r-matrix formulation
associates classical integrable systems with loop algebras. In the case of the classical Gaudin models, given a simple Lie algebra $\alg$, one can define three different kinds of $N$-body integrable sistems on the tensor product of $N$ copies of the corresponding Lie--Poisson algebra $\cal{F}(\alg^*)$: rational, trigonometric and elliptic Gaudin models. To these integrable models correspond the rational, trigonometric and 
elliptic Gaudin algebras, respectively. If one considers the homogeneous rational Gaudin models, it turns out that some of its Hamiltonians are superintegrable, since they belong also to a set of involutive functions obtained through the so called ``coproduct method'' \cite{Ragn,Balrag,CRMAngel} and \cite{Vulpi} for a review.
The superintegrable Gaudin Hamiltonians
are obtained when one applies the method using the standard coalgebra structure defined on a Lie-Poisson algebra thus establishing the relation
between the coproduct method and the Gaudin models (see \cite{musso}). However, the second set of involutive functions generated through the
coproduct method is complete only when  $\alg=\mathfrak{sl}(2)$. The problem of completing this set of functions has been solved when $\alg=\mathfrak{sl}(N)$ in \cite{FM} using a bihamiltonian formulation.

In the present paper we propose a generalisation of both the coproduct method and of the Gaudin algebras, that we call ``loop coproduct method''. 
This generalisation works only in the classical case, but allows us to define a larger number of involutive functions with respect to the standard coproduct method and,
in particular, to recover all the missing integrals of the standard coproduct method when $\alg$ is a simple Lie algebra.
Moreover, rational, trigonometric and elliptic Gaudin models can be obtained as particular cases of this construction, so that it gives
a unified theoretical framework for these algebraic approaches to classical integrable models. We also show how, through this generalisation, 
it is possible to treat in a natural way the rational Gaudin model with degeneracies and that, in this context, the coproduct method associated with the standard coalgebra structure of the Lie-Poisson algebras correspond to the case of maximal degeneracy. 

The plan of the paper is the following. In section \ref{coproduct} we briefly recall the coproduct method. In section \ref{Gaudinal} we recall
the definition of the Gaudin models and of the Gaudin algebras. The loop-coproduct method is introduced in section \ref{loopcoproduct}, where we
also show that both the coproduct method and the Gaudin algebras are particular cases of this approach. 
In section \ref{copGaudin} we show how rational Gaudin models with degeneracies can be obtained through the loop-coproduct method.
Section \ref{higher} shows how applying this approach to Lie--Poisson algebras one can obtain a larger set of involutive functions
with respect to the coproduct method. We derive in subsection \ref{simpleal} that when the Lie--Poisson algebra is simple these functions do indeed
define a completely integrable system and in subsection \ref{twophoton} we give an explicit example for a non-simple algebra. 
Finally in section \ref{conclusion} we discuss the results we have presented and we comment on open problems.

\section{Coproduct method} \label{coproduct}

The coproduct method is a recipe to construct (classical and quantum) Hamiltonian integrable systems endowed with Poisson-coalgebra symmetry. 
In this section we tersely recall how it works in the classical case only, for detailed expositions, including different applications, 
we refer the reader to \cite{Ragn,Balrag,CRMAngel,Vulpi,alfonso,alfonsoh6}. 

Let $A$ be a unital, associative algebra and let us denote its identity map with $id$.
If $A$ is also endowed with a coproduct map
$$
\Delta: A
\rightarrow A\otimes A
$$
which is  coassociative
$$
(\Delta \otimes id) \circ \Delta=(id \otimes \Delta) \circ \Delta
\label{coas}
$$
and is an  algebra morphism from
$A$ to $A\otimes A$:
$$
\Delta (a\,b)=\Delta (a)\,\Delta (b) ,\qquad \forall\, a,b \in A,
$$
then $(A,\Delta)$ is called a coalgebra.
If $A$ is a Poisson algebra and $\Delta$ is also a Poisson morphism:
$$
\Delta(\pois{a}{b}_A)=\pois{\Delta(a)}{\Delta(b)}_{A\otimes A} ,\qquad \forall
a,b \in A ,
$$
with respect to the standard poisson structure on $A \otimes A$:
$$
\pois{a \otimes b}{c \otimes d}_{A \otimes A}\doteq \pois{a}{c}_A \otimes b d + a c \otimes \pois{b}{d}_A \qquad
a,b,c,d \in A
$$
we shall say that $(A,\Delta)$ is a  Poisson coalgebra.
Let $M$ be the dimension of $A$, $\{y^\alpha\}_{\alpha=1}^M$ a set of generators and let us suppose that
$r$ functionally independent Casimir functions ${\cal{C}}_j(\vec{y}) \equiv {\cal{C}}_j(y^1,\dots,y^M)$ are defined on $A$. 
Using the coproduct and the identity maps it is possible to construct the $m$-th coproduct maps: 
\begin{eqnarray}
&& \co^{(m)}:A\rightarrow  \overbrace{A\otimes A \otimes \dots \otimes A}^m \qquad m=2,\dots,N \nonumber\\
&& \co^{(m)}:=(\co^{(m-1)} \otimes id)\circ\co^{(2)} \qquad \qquad \co^{(2)}\doteq \Delta \label{mth}
\end{eqnarray}
that, in turn, are also both algebra and Poisson morphisms:
\begin{eqnarray*}
&& \co^{(m)}(a\,b)=\co^{(m)}(a)\,\co^{(m)} (b) ,\qquad \forall\, a,b \in A,\\
&& \co^{(m)}(\pois{a}{b}_A)=\pois{ \co^{(m)}(a)}{ \co^{(m)}(b)}_{A\otimes
A\otimes\dots^{m)}\otimes A } \qquad \forall
a,b \in A
\end{eqnarray*} 
Let $a$ be an element of $\overbrace{A\otimes A \otimes \dots \otimes A}^m$ and $b$ be an element of $\overbrace{A\otimes A \otimes \dots \otimes A}^n$.
Then we define 
$$
\left\{ a, b \right\}_{A\otimes
A\otimes\dots^{N)}\otimes A} \doteq \{ a \otimes \overbrace{1\otimes 1 \otimes \dots \otimes 1}^{N-m}, b \otimes \overbrace{1\otimes 1 \otimes \dots \otimes 1}^{N-n} \}_{A\otimes
A\otimes\dots^{N)}\otimes A}
$$
In the following all the Poisson brackets will be considered in $\overbrace{A\otimes A \otimes \dots \otimes A}^N$, so that the indication of the space in the Poisson bracket will be dropped.   

From the Poisson morphism property of the $m-$th coproduct it follows:
\begin{equation}
\pois{\co^{(m)}({\cal{C}}_i)}{{\co^{(n)}(y^\alpha)}}=0 \qquad {\rm if} \ n \geq m. \label{fund}
\end{equation}
From equation (\ref{fund}) it follows at once that:
\begin{eqnarray}
&& \back \pois{\co^{(m)}({\cal{C}}_i)}{\co^{(n)}({\cal{C}}_j)}=0, \qquad  m,n=2,\dots,N, \quad
i,j=1,\dots,r \label{fund2a} \\
&& \back \pois{\co^{(m)}({\cal{C}}_i)}{\co^{(N)}(y^\alpha)}=0, \qquad  m=2,\dots,N, \quad
i=1,\dots,r \quad \alpha=1,\dots,M \label{fund2b}
\end{eqnarray}
Hence, given a Poisson coalgebra $A$, the coproduct method allow us to find a large family of functions in involution within the Poisson 
algebra $\overbrace{A\otimes A \otimes \dots \otimes A}^N$. 

In the following we will work mainly with the particular case when $A$ is a Lie-Poisson algebra ${\cal{F}}(\alg^*)$ with commutation relations
$$
\{y^\al, y^\be \}= C^{\al \be}_\ga \, y^\ga
$$ 
and the
natural coalgebra structure
$$
\Delta(y^\al)=y^\al \otimes 1 + 1 \otimes y^\al \equiv y^\al_1+ y^\al_2.
$$ 
From now on, we will denote with $\{y^\al_i\},\ \al=1,\dots,M, \ i=1,\dots,N$ the natural basis in 
$\overbrace{A \otimes \dots \otimes A}^N$:
\begin{eqnarray*}
&& y^\al_i=\overbrace{1 \otimes \dots \otimes 1}^{i-1} \otimes y^\alpha \otimes \overbrace{1 \otimes \dots \otimes 1}^{N-i}\\ 
&& \{y^\al_i,y^\be_j\}=\delta_{ij} \ C^{\al \be}_\ga y^\ga_i.
\end{eqnarray*}  
The m-th coproducts can be obtained  from equation (\ref{mth}) and in this basis they simply read
\begin{equation}
\Delta^{(m)}(y^\al)=\sum_{i=1}^m y^\al_i, \label{stdmth}
\end{equation}
then it is immediate to check that
\begin{equation}
\left\{ \Delta^{(i)}(y^\al), \Delta^{(j)}(y^\be) \right\}= C^{\al \be}_\ga \, \Delta^{(i)}(y^\ga) \qquad j\geq i. \label{standard}
\end{equation}

Let us now go back to the general case of $A$ being a Poisson algebra with $r$ Casimirs function. It is natural to ask when the family of involutive functions obtained through the coproduct method will be large enough to define an integrable system.
The dimension of the symplectic leaves of $A$ is:
$$
d=\frac{M-r}{2}, 
$$ 
so the total number of degrees of freedom of an Hamiltonian system defined on $\overbrace{A \otimes \dots \otimes A}^N$ is $N(M-r)/2$.
Equation (\ref{fund2a}) defines a family of $(N-1)r$ involutive functions, while equation (\ref{fund2b}) allow us to 
add to this family a maximal abelian subalgebra of $\Delta^{(N)}(A) \simeq A$; that is at most $(M-r)/2$ further
involutive functions.
Hence, the coproduct method allow us to define an involutive family of at most
$$
(N-1)r + \frac{M-r}{2}
$$
independent functions.
Henceforth, a necessary condition for getting a complete integrable Hamiltonian system through the coproduct method is given by: 
\begin{equation}
(N-1)r + \frac{M-r}{2} \geq N \, \frac{M-r}{2} \quad
\Longrightarrow \quad 2 r \geq M-r, \label{conto} 
\end{equation}
and this condition is almost never satisfied for Poisson coalgebras. For example, if $A$ is a Lie--Poisson coalgebra constructed out from a simple Lie algebra $\alg$, then the condition (\ref{conto}) is satisfied only when $\alg=sl(2)$.

A natural question, when the coproduct method do not grant complete integrability, is if it is possible to enlarge in some way the family of integrals
in order to define an integrable system.
In section \ref{simpleal} we show that, at least in the case of simple Lie--Poisson coalgebras, the answer to this question is affirmative. 

\section{Gaudin models and Gaudin algebras} \label{Gaudinal}

In the modern formulation,
$N$-body classical Gaudin models are integrable models defined on the tensor product of $N$ copies of a simple Lie-Poisson algebra 
$\cal{F}(\alg^*)$. Indeed, given a simple Lie algebra $\alg$ and a solution $r(\la) \in \alg \otimes \alg$ of the Classical Yang-Baxter equation: 
\begin{equation}
[r_{13}(\la),r_{23}(\mu)]+[r_{12}(\la-\mu),r_{13}(\la)+r_{23}(\mu)]=0 ,\label{cYB}
\end{equation}
where 
$$
r_{12}(\la) = r(\la)  \otimes \mathds{1}, \qquad r_{23}(\la)=\mathds{1} \otimes r(\la), \qquad
r_{13}(\la)= ( \mathds{1} \otimes \Pi)\, r_{12}(\la) \, ( \mathds{1} \otimes \Pi)
$$
and $\Pi$ is the permutation operator on $\alg \otimes \alg$:
$$
\Pi(x \otimes y)= y \otimes x,
$$
there exists a simple recipe (see, for example \cite{MPR} or \cite{Skrypnyk}) to construct a Lax matrix $L(\la)$ with entries in 
$\cal{F}(\alg^*)^{\otimes \, N}$ together with an $r$-matrix $r'(\la)$ satisfying:
\begin{equation}
\{ L(\la) \otimes \mathds{1}, \mathds{1} \otimes L(\mu) \}+[r'(\la-\mu), 
 L(\la) \otimes \mathds{1}+\mathds{1} \otimes L(\mu)]=0. \label{BV}
\end{equation} 
Equation (\ref{BV}) ensures the commutativity of the spectral invariants of the Lax matrix $L(\la)$ that give rise to the Gaudin Hamiltonians. 
Drinfel'd and Belavin \cite{D}, proved that under suitable non-degeneracy conditions on $r(\la)$,
the dependency on the spectral parameter can be only of three kinds: rational, trigonometric and elliptic.
According to the chosen solution, equation (\ref{BV}) will define the rational, trigonometric or elliptic Gaudin algebra (actually elliptic Gaudin algebras can be defined only for $\alg=\mathfrak{sl}(n)$). 

Let us make this construction explicit.
Let $\rho$ be a faithful representation of $\alg$, then the trace define a nondegenerate invariant bilinear form on $\rho(\alg)$. Let 
$\{X^\al\}_{\al=1}^M$, where $M$ is the dimension of $\alg$, an orthonormal basis with respect to the trace, which means that in this basis the Cartan-Killing metric is diagonal\footnote{For the sake of keeping the notation simple and make contact with the one used in \cite{D} we decided
to work in an orthonormal basis, hence considering $\alg$ as a complex Lie algebra. The construction works ``mutatis mutandis'' also for 
the real forms of $\alg$ and for an arbitrary choice of the basis (see \cite{Skrypnyk}).} 
\begin{equation}
g^{\al \be}={\rm Tr}(X^\al X^\be)=\delta^{\al \be} \label{ortho}
\end{equation}
and the corresponding structure constants
$$
[X^\alpha, X^\be]=C^{\alpha \beta}_\gamma X^\ga  
$$
are completely antisymmetric:
$$
C^{\alpha \beta}_\gamma=-C^{\alpha \gamma}_\beta
$$

In \cite{D} it has been proven that the classical Yang-Baxter equation (\ref{cYB}) always admits matrix solutions of the form
(from now on we will write explicitly the sum in case Einstein convention could generate confusion):
\begin{equation}
r'(\la)=\sum_{\al=1}^M X^\al \otimes X^\al f^\al(\la) \label{rprimo}
\end{equation} 
where $f^\al(\la)$ are suitable functions that can have an elliptic, trigonometric or rational dependence on the spectral parameter $\la$.

Let $\{y^\alpha_j\},\alpha=1,\dots,M$ be the coordinate functions on the Lie--Poisson algebra ${\cal{F}}(\alg^*)^{\otimes N}$
corresponding to the basis $\{X^\al\}_{\al=1}^M$. The ``fundamental'' Poisson brackets among the coordinates will then be given by:
$$
\{y^\alpha_i,y^\be_j \}=\delta_{ij} C^{\alpha \beta}_\gamma y^\ga_i.
$$ 
The Lax matrix of the $N-$body Gaudin model associated with the r-matrix (\ref{rprimo}) is given by:
\begin{equation}
L(\la)=\sum_{\alpha=1}^M \sum_{j=1}^N  X^\alpha f^\al(\la-\ep_j) y^\al_j +\sigma, \label{Lgau}
\end{equation}
where $\ep_j$ are constant parameters subject to the constraint $\ep_i \neq \ep_j,\ i \neq j$, the functions $f^\al(\la)$ are the same 
appearing in (\ref{rprimo}) and $\sigma$ is a constant matrix satisfying the constraint
\begin{equation}
[\sigma \otimes 1 + 1 \otimes \sigma, r(\la)]=0. \label{cond2}
\end{equation}
It can be proven (see for example \cite{MPR} or \cite{Skrypnyk}) that the Gaudin Lax matrix (\ref{Lgau}) satisfies the $r-$matrix formulation (\ref{BV}) with the $r-$matrix (\ref{rprimo}).

The Gaudin algebra generators are defined by
\begin{equation}
y^\al(\la)={\rm Tr} (L(\la) X^\al)=\sum_{j=1}^N f^\al(\la-\ep_j) y^\al_j + {\rm Tr}(\sigma X^\al) \label{Gaugen}
\end{equation}
Let us evaluate the Poisson brackets between the Gaudin algebra generators (\ref{Gaugen}):
\begin{eqnarray*}
&& \{y^\al(\la), y^\be(\mu)\}={\rm Tr} \left(\{ L(\la) \otimes \mathds{1}, \mathds{1} \otimes L(\mu) \} X^\al \otimes X^\be \right)=\\
&& =-{\rm Tr} \left( [r(\la-\mu), 
 L(\la) \otimes \mathds{1}+\mathds{1} \otimes L(\mu)] X^\al  \otimes X^\be \right)=\\
&&=-{\rm Tr} \left( \left[ \sum_\ga (X^\ga \otimes X^\ga) f^\ga(\la-\mu), \sum_\delta (X^\de \otimes 1) 
y^\de(\la)+\right. \right.\\
&& \left. \left. +  \sum_\delta (1 \otimes X^\de) y^\de(\mu) \right] (X^\al \otimes X^\be) \right) =\\
&&= - \sum_{\ga\de} f^\ga(\la- \mu) C^{\ga \de}_\ep {\rm Tr} \left( \left((X^\ep \otimes X^\ga) y^\de(\la)+(X^\ga \otimes X^\ep) y^\de(\mu) \right) X^\al \otimes X^\be \right) =\\
&&=  - \sum_{\ga\de\ep} f^\ga(\la- \mu) C^{\ga \de}_\ep \left(
\delta^{\al \ep} \delta^{\ga \be} y^\de(\la) +\delta^{\al \ga} \de^{\be \ep} y^\de(\mu) 
\right)=\\
&&= -\sum_{\de} \left(f^\be(\la-\mu) C^{\be \de}_\al  y^\de(\la) + f^\al(\la-\mu) C^{\al \de}_\be y^\de(\mu) \right).  
\end{eqnarray*}
By using the fact that the structure constants are completely antisymmetric, finally we get:
\begin{eqnarray}
&& \{y^\al(\la), y^\be(\mu)\}=-\sum_{\de} \left( f^\be(\la-\mu) C^{\be \de}_\al  y^\de(\la) + f^\al(\la-\mu) C^{\al \de}_\be y^\de(\mu) \right) = \nonumber \\
&&=\sum_\de C^{\al \be}_\de \left(f^\al(\la-\mu)  y^\de(\mu)-  f^\be(\la-\mu) y^\de(\la) \right). \label{Gaugen2}
\end{eqnarray}
These are the defining relations of the Gaudin algebra associated with $\alg$ and $r'$ in the orthonormal basis (\ref{ortho}).

\subsection{The rational Gaudin model} \label{subsec}
In the following we will deal mainly with the rational case:
$$
f^\al(\la)=\frac{1}{\la} \qquad  \alpha=1,\dots,M. 
$$
Accordingly, the Lax matrix of the rational Gaudin model associated with $\alg$ is given by
\begin{equation}
L(\la)=\sum_{\al=1}^M \sum_{i=1}^N \frac{ X^\al y^\al_i}{\la-\epsilon_i}+ \sigma \label{Lax}.
\end{equation} 
In the rational case equation (\ref{cond2}) is always satisfied when $\sigma$ belongs to $\rho(\alg)$.
The rational Gaudin algebra generators are given by:
\begin{equation}
y^\al(\la)={\rm Tr}(L(\la) X^\al)=\sum_{i=1}^N \frac{y^\al_i}{\la-\epsilon_i}+ {\rm Tr}(\sigma X^\al)= \sum_{i=1}^N \frac{y^\al_i}{\la-\epsilon_i}+
c^\al \label{ydef}
\end{equation} 
and the Gaudin algebra by the Poisson brackets:
\begin{equation}
\{y^\al(\la), y^\be(\mu) \}=\frac{1}{\mu-\la} C^{\al \be}_\ga(y^\ga(\la)-y^\ga(\mu)). \label{Gaudinalgebra}
\end{equation}
Let $r={\rm rank} \, \alg$ and let us denote with ${\cal{C}}_1, \dots, {\cal{C}}_r$ the fundamental Casimir functions in ${\cal{F}}(\alg^*)$.
The rational Gaudin Hamiltonians are given by the spectral invariants of the Lax matrix (\ref{Lax}) or, which is the same, by the residues of the Casimir functions evaluated on the rational Gaudin algebra generators (\ref{ydef}):
$$
{\rm Tr} \left( L(\la)^{m_j+1} \right) = {\cal{C}}_j(y^1(\la), \dots, y^M(\la))
$$
where $m_j$, $j=1,\dots,r$ are the ``exponents'' \cite{Fuchs} of the Lie algebra $\alg$. 
When $\sigma$ has simple spectrum, the Gaudin Hamiltonians define a completely integrable system (provided that the parameters $\epsilon_i$ are
all distinct) \cite{RSTS}. If this is not the case, then there are missing integrals. These integrals can be recovered noticing that if $\tau$ is 
a matrix belonging to the commutant of $\sigma$, i.e. $[\sigma,\tau]=0$ then the function:
\begin{displaymath}
F_\tau=\sum_{\alpha=1}^M \sum_{i=1}^N  y^\al_i {\rm Tr}  \left(  X^\al \tau \right)
\end{displaymath}
commutes with all the spectral invariants of (\ref{Lax}). By adding to the spectral invariants of (\ref{Lax}) a maximal abelian subalgebra of the Poisson algebra spanned by such functions
\begin{equation}
\{ F_\tau=\sum_{\alpha=1}^M \sum_{i=1}^N  y^\al_i {\rm Tr} \left(  X^\al \tau \right) \quad | \quad \ [\tau,\sigma]=0 \},   \label{Ftau}  
\end{equation}
complete integrability is recovered.

When $\sigma=0$, the algebra  (\ref{Ftau}) coincide with the whole algebra $\Delta^{(N)}({\cal F}(\alg^*))$,  
where $\Delta^{(N)}$ denotes the primitive $N$-th coproduct
map associated with the Lie-Poisson algebra ${\cal F}(\alg^*)$ (see equation \ref{stdmth}).
In such case, some of the rational Gaudin Hamiltonians turn out to be superintegrable. Indeed, it can be shown that 
$\Delta^{(N)}({\cal{C}}_1), \dots, \Delta^{(N)}({\cal{C}}_r)$, belong to the Gaudin Hamiltonians \cite{musso}. 
Therefore, these Hamiltonians belong
to two different sets of functions in involution, the one obtained from the spectral invariants of the Lax matrix (\ref{Lax}) and the
one generated through the coproduct method (see equations \ref{fund2a} and \ref{fund2b}).

\section{Loop coproducts} \label{loopcoproduct}

Let $A$ be a generic Poisson algebra and $\{y^\alpha\}_{\alpha=1}^M$ a set of generators for $A$ with Poisson brackets:
$$
\{ y^\al, y^\be \}=F^{\al \be}(\vec{y}) \qquad \vec{y}=(y^1,\dots,y^M). 
$$
Let us denote with $\vec{Y}$ the set of all the generators on $A^{\otimes N}$:
\begin{eqnarray*}
&& \vec{Y} \equiv \{y^\al_i\} \qquad \al=1,\dots,M \quad i=1,\dots,N\\
&& y^\al_i \equiv \overbrace{1 \otimes \dots \otimes 1}^{i-1} \otimes \, y^\al \otimes \overbrace{1 \otimes \dots \otimes 1}^{N-i} 
\end{eqnarray*}
We state our main result:
\begin{theorem}{\label{main}}
Let us suppose that we have defined a set of $m$ maps depending on a parameter $\la$
\begin{equation}
\Delta^{(k)}_{\lambda}: A \to \overbrace{A \otimes \dots A \otimes A}^{N}, \quad k=1, \dots, m \label{lcopros}
\end{equation}
and such that the following relations hold
\begin{eqnarray}
&&  \{ \Delta_{\la}^{(i)}(y^\al), \Delta_{\mu}^{(k)}(y^\beta) \}=f^\beta_\gamma(i,k,\la,\mu,\vec{Y}) F^{\al \ga} (\Delta_{\la}^{(i)}(\vec{y})) \quad k>i \label{p1}\\
&& \{ \Delta_{\la}^{(k)}(y^\al), \Delta_{\mu}^{(k)}(y^\be) \}= \nonumber \\
&& = g^\beta_\ga(k,\la,\mu,\vec{Y}) F^{\al \ga} (\Delta_{\la}^{(k)}(\vec{y}))+ h^\al_\ga(k,\la,\mu,\vec{Y}) F^{\ga \be} (\Delta_{\mu}^{(k)}(\vec{y}))\label{p2}
\end{eqnarray}
for certain functions $f^\beta_\ga(i,k,\la,\mu,\vec{Y}),g^\beta_\ga(k,\la,\mu,\vec{Y}),h^\al_\ga(k,\la,\mu,\vec{Y})$.

If the map $\Delta_{\la}^{(i)}$ is defined on any smooth function of the generators $f \in A$ as: 
$$
\Delta^{(i)}_{\lambda}(f)(y^1,\dots,y^M)) = f(\Delta^{(i)}_{\lambda}(y^1),\dots,\Delta^{(i)}_{\lambda}(y^M)),
$$
then:
\begin{eqnarray}
 \{ \Delta^{(i)}_{\lambda}({\cal{C}}_j) ,  \Delta^{(k)}_\mu (y^\be) \}&=&0 \quad k > i  \label{lambda1} \\
 \{ \Delta^{(i)}_{\lambda}({\cal{C}}_j) ,  \Delta^{(k)}_{\mu}({\cal{C}}_l) \}&=&0.  \label{lambda2}
\end{eqnarray} 
\end{theorem}  

\noindent {\bf Proof:} 

\noindent Let us prove first the equation (\ref{lambda1}). We fix an arbitrary Casimir ${\cal{C}}_j$.
Since ${\cal{C}}_j$ is a Casimir function, for any $\beta$ we must have:
\begin{eqnarray}
\pois{{\cal{C}}_j}{y^\be}=\sum_{\al=1}^M \part{{\cal{C}}_j}{y^\al} \pois{y^\al}{y^\be}=\sum_{\al=1}^M \part{{\cal{C}}_j}{y^\al} 
F^{\al \be}(\vec{y}) =0  \label{classic} 
\end{eqnarray}
Now we use this equation together with (\ref{p1}) to prove equation (\ref{lambda1}):
\begin{eqnarray*}
&&  \{ \Delta^{(i)}_{\lambda}({\cal{C}}_j),  \Delta^{(k)}_\mu (y^\be) \}=\sum_{\al=1}^{M}  \part{{\cal{C}}_j \left(\Delta^{(i)}_{\lambda}(\vec{y}) \right)}{\Delta^{(i)}_\la (y^\al)} \,
\pois{\Delta^{(i)}_\la (y^\al)}{\Delta^{(k)}_\mu (y^\be) }=\\  
&&  =f^\beta_\gamma(i,k,\la,\mu,\vec{Y}) \sum_{\al=1}^{M} \part{{\cal{C}}_j \left(\Delta^{(i)}_{\lambda}(\vec{y}) \right)}{\Delta^{(i)}_\la (y^\al)} \,  F^{\al \ga} (\Delta_{\la}^{(i)}(\vec{y}))=0   \qquad k>i
\end{eqnarray*}
From equation 
$$
 \{ \Delta^{(i)}_{\lambda}({\cal{C}}_j) ,  \Delta^{(k)}_\mu(y^\al) \}=0 \quad k > i
$$
it follows that 
$$
\{\Delta^{(i)}_{\lambda}({\cal{C}}_j),  \Delta^{(k)}_{\mu}({\cal{C}}_l) \}=0 \qquad i \neq k
$$
so that only the case $k=i$ remains to be proven. In this case:
\begin{eqnarray*}
&&  \{ \Delta^{(i)}_{\lambda}({\cal{C}}_j) ,  \Delta^{(i)}_{\mu}({\cal{C}}_l) \}=\\
&&  = \sum_{\al,\be=1}^M \part{{\cal{C}}_j \left(\Delta^{(i)}_{\lambda}(\vec{y}) \right)}{\Delta^{(i)}_\la (y^\al)}   \part{{\cal{C}}_l \left(\Delta^{(i)}_{\mu}(\vec{y}) \right)}{\Delta^{(i)}_\mu (y^\be)} \,\pois{\Delta^{(i)}_\la (y^\al)}{\Delta^{(i)}_\mu (y^\be) }
\end{eqnarray*} 
By using equation (\ref{p2}) and (\ref{classic}), we have:
\begin{eqnarray*}
&&  \{ \Delta^{(i)}_{\lambda}({\cal{C}}_j) ,  \Delta^{(i)}_{\mu}({\cal{C}}_l) \}=\\
&&  = \sum_{\be} \part{{\cal{C}}_l \left(\Delta^{(i)}_{\mu}(\vec{y}) \right)}{\Delta^{(i)}_\mu (y^\be)} 
g^\beta_\ga(k,\la,\mu,\vec{Y}) \left( \sum_{\al=1}^{M}  \part{{\cal{C}}_j \left( \Delta^{(i)}_{\lambda}(\vec{y})\right)}{\Delta^{(i)}_\la (y^\al)} F^{\al \be}( \Delta_{\la}^{(i)}(\vec{y})) \right) + \\
&&  + \sum_{\al} \part{{\cal{C}}_j \left(\Delta^{(i)}_{\la}(\vec{y}) \right)}{\Delta^{(i)}_\la (y^\al)} 
h^\al_\ga(k,\la,\mu,\vec{Y}) \left( \sum_{\be=1}^{M}  \part{{\cal{C}}_l \left( \Delta^{(i)}_{\mu}(\vec{y})\right)}{\Delta^{(i)}_\mu (y^\be)} F^{\ga \be}( \Delta_{\mu}^{(i)}(\vec{y})) \right)=0
\end{eqnarray*} 
since the terms in parentheses vanish, which completes the proof. 
\endpf

Hereafter we call the maps (\ref{lcopros}) as ``loop coproducts'' and the algebra (\ref{p1})-(\ref{p2}) as ``generalized Gaudin algebra''. In fact, 
it is straightforward to realize that
the generators of the Gaudin algebras  (see eq. (\ref{Gaugen2})) are particular cases of (\ref{p1}), (\ref{p2}) provided that 
\begin{eqnarray*}
g^\beta_\ga(k,\la,\mu,\vec{Y})=-\delta^{\be}_\ga f^\be(\la-\mu)\\
h^\al_\ga(k,\la,\mu,\vec{Y})=\delta^{\al}_\ga f^\al(\la-\mu).
\end{eqnarray*}

Also the coproduct construction is a particular case of Theorem \ref{main}, since the $m-$th coproduct
map satisfy (\ref{p1}) and (\ref{p2}) for suitable choices of the arbitrary functions $f^\beta_\ga(i,k,\la,\mu,\vec{Y})$,$g^\beta_\ga(k,\la,\mu,\vec{Y})$ and $h^\beta_\ga(k,\la,\mu,\vec{Y})$. 

Let us show it.
We recall, from section \ref{coproduct} that we have to consider the case of $A$ being a Poisson coalgebra. Generically, the coproduct of a generator can be written as the finite sum (see \cite{Balrag})
\begin{equation}
\Delta^{(2)}(y^\al)=\sum_b G^\al_b(\vec{y}) \otimes H^\al_b(\vec{y}) \label{defco1}
\end{equation}  
for some functions $G^\al_b, \ H^\al_b$. 
As shown in \cite{Balrag}, thanks to the coassociativity property of the coproduct, the $k-$th coproduct map can be written as
\begin{equation}
\Delta^{(k)}=\left( \Delta^{(i)} \otimes \Delta^{(k-i)} \right) \circ \Delta^{(2)} \qquad i<k \label{defco2}
\end{equation}
So that, putting together (\ref{defco1}) and (\ref{defco2}), we have
\begin{equation}
\Delta^{(k)}(y^\al)=\sum_b G^\al_b(\Delta^{(i)}(\vec{y})) \otimes H^\al_b(\Delta^{(k-i)}(\vec{y})) \qquad i<k. \label{defco3}
\end{equation} 
Let us now assume that $i<k$, by using (\ref{defco3}) we have:
\begin{eqnarray*}
&&   \{ \Delta^{(i)}(y^\al), \Delta^{(k)}(y^\beta) \}=\sum_b \left[ \{ \Delta^{(i)}(y^\al),
G^\be_b(\Delta^{(i)}(\vec{y})) \} \otimes H^\be_b(\Delta^{(k-i)}(\vec{y})) \right] = \\
&&  = \sum_b \left[ \left( \part{G^\be_b(\Delta^{(i)}(\vec{y}))}{\Delta^{(i)} (y^\ga)} \{ \Delta^{(i)}(y^\al), \Delta^{(i)}(y^\ga) \} \right)
\otimes H^\be_b(\Delta^{(k-i)}(\vec{y})) \right]=\\
&&  = \sum_b \left[ \left( \part{G^\be_b(\Delta^{(i)}(\vec{y}))}{\Delta^{(i)} (y^\ga)} F^{\al \ga}(\Delta^{(i)}(\vec{y}))  \right)
\otimes H^\be_b(\Delta^{(k-i)}(\vec{y})) \right]
\end{eqnarray*}
which is just of the form (\ref{p1}) with 
$$
f^\be_\ga(i,k,\la,\mu,\vec{Y})= \sum_b \part{G^\be_b(\Delta^{(i)}(\vec{y}))}{\Delta^{(i)} (y^\ga)} \otimes H^\be_b(\Delta^{(k-i)}(\vec{y}))
$$
When $k=i$, from the homomorphism property of the coproduct, we simply have
$$
\{ \Delta^{(i)}(y^\al), \Delta^{(i)}(y^\beta) \}=F^{\al \be}(\Delta^{(i)}(\vec{y}))  
$$
which is of the form (\ref{p2}) with (for example):
$$
g^\beta_\ga(k,\la,\mu,\vec{Y})= \delta^\be_\ga \qquad  h^\beta_\ga(k,\la,\mu,\vec{Y})=0.
$$
Therefore, the loop coproduct construction can be interpreted as a simultaneous generalization of both the coproduct method and of the Gaudin
algebra approach.

\section{Rational Gaudin model with degeneracies} \label{copGaudin}

In this section we will show how the case of the rational Gaudin model with degeneracies can be naturally formulated in the loop coproduct scheme.

Let $\alg$ be a simple Lie algebra as in section \ref{Gaudinal} and $A={\cal F}(\alg^*)$. 
In the Subsection \ref{subsec}, we have seen (see eq. (\ref{ydef})) that the map 
$$
\Delta_{\lambda}: A \to \overbrace{A \otimes \dots A \otimes A}^{N}
$$
given by:
\begin{equation}
\Delta_{\lambda}(y^\al) \equiv y^\al(\la)= \sum_{i=1}^N \frac{y^\al_i}{\la-\epsilon_i} +c^\al \label{loopcopGau}
\end{equation}
with $c^\al, \epsilon_1, \dots, \epsilon_N$ arbitrary constant parameters, defines the rational Gaudin algebra (\ref{Gaudinalgebra}):
$$
\{ \Delta_{\lambda}(y^\al), \Delta_{\mu}(y^\be) \} = \frac{1}{\mu-\la} C^{\al \be}_\ga  \left( \Delta_{\lambda}(y^\ga)- \Delta_{\mu}(y^\ga)\right) 
$$
Hence $\Delta_{\lambda}$ satisfies the property (\ref{p2}) with 
$$
g(\la,\mu)=-h(\la,\mu)=\frac{1}{\mu-\la}
$$
and the residues of the images of the Casimir functions under $\Delta_{\lambda}$, $\Delta_{\lambda}({\cal C}_i), \ i=1,\dots,r$ define the 
rational $\alg$-Gaudin Hamiltonians: 
\begin{equation}
\pois{\Delta_{\lambda}({\cal{C}}_i)}{\Delta_{\mu}({\cal{C}}_j)}=0 \qquad i,j=1,\dots,r \label{Gaudin}
\end{equation}
If the parameters $\epsilon_1,\dots,\epsilon_N$ are all distinct: $\epsilon_i \neq \ep_j$, for $i \neq j$ and 
the constant matrix 
$$
\sigma=\sum_{\al} X^\al c^\al
$$
has simple spectrum, then 
equation (\ref{Gaudin}) provides enough functions in involution to get complete integrability. 

In the Subsection \ref{subsec} we briefly discussed what happens when the spectrum of $\sigma$ is not simple. In this Section we
will assume that $\sigma$ has simple spectrum and we will discuss the case  
when there are degeneracies in the parameters, i.e. $\ep_i=\ep_j$ for some $i \neq j$. Also in this case, the spectral invariants of
the Lax matrix (\ref{Lax}) are not enough to define a completely integrabile system.
However, the Gaudin model with degeneracies is still integrable, and the expressions of the missing integrals was given in 
\cite{Kuznetsov} for the $\alg=sl(2)$ case, and recently in \cite{Falqui} for the case of an arbitrary simple Lie--Poisson algebra. 
In the following we show that such missing integrals can be recovered through the loop coproduct formulation.  
Let us assume that we have $k$ distinct values for the parameters $\ep_i$, each with multiplicity $d_i,\ i=1,\dots,k$. By renumbering the copies in $A^{\otimes N}$ and the $\ep$ parameters, we can put
\begin{equation}
\begin{array}{l}
\ep_1=\ep_2=\dots=\ep_{d_1} =\delta_1 \\
\ep_{d_1+1}=\ep_{d_1+2}=\dots=\ep_{d_1+d_2}=\delta_2 \\
\vdots   \\
\ep_{N-d_k+1}=\ep_{N-d_k+2}=\dots=\ep_{N}=\delta_k
\end{array}
\end{equation} 
Let us introduce the $k+1$ maps:
\begin{eqnarray*}
&& \Delta^{(1)}_\la(y^\al)=\sum_{i=1}^{d_1} \frac{y^\al_i}{\la-\eta_i} \\
&& \Delta^{(2)}_\la(y^\al)=\sum_{i=d_1+1}^{d_2} \frac{y^\al_i}{\la-\eta_i} \\
&& \vdots \\
&&  \Delta^{(k)}_\la(y^\al)=\sum_{i=N-d_k+1}^{N} \frac{y^\al_i}{\la-\eta_i} \\
&& \Delta^{(k+1)}_\la(y^\al) \equiv \Delta_\la (y^\al)=\sum_{i=1}^{d_1} \frac{y^\al_i}{\la-\delta_1}+ \sum_{i=d_1+1}^{d_2} \frac{y^\al_i}{\la-\delta_2}+
\dots + \sum_{i=N-d_k+1}^{N} \frac{y^\al_i}{\la-\delta_k} +c^\al\\
\end{eqnarray*}
where the $\eta_i$ are $N$ distinct parameters such that $\eta_i \neq \eta_j, \ i\neq j$.  
These maps satisfy:
\begin{eqnarray*}
&& \pois{\Delta^{(i)}_\la(y^\al)}{\Delta^{(j)}_\mu(y^\be)}=0 \qquad i,j=1,\dots,k\\
&& \pois{\Delta^{(i)}_\la(y^\al)}{\Delta^{(k+1)}_\mu(y^\be)}= \frac{1}{\mu-\delta_{i}} C^{\al \be}_\ga  \, \Delta^{(i)}_{\lambda}(y^\ga) \qquad i=1,\dots,k\\
&& \{ \Delta^{(k+1)}_{\lambda}(y^\al), \Delta^{(k+1)}_{\mu}(y^\be) \} = \frac{1}{\mu-\la} C^{\al \be}_\ga  \left( \Delta^{(k+1)}_{\lambda}(y^\ga)- \Delta^{(k+1)}_{\mu}(y^\ga)\right). 
\end{eqnarray*}
By Theorem 1 we have that
$$
\{ \Delta^{(i)}_{\lambda}({\cal{C}}_l) ,  \Delta^{(j)}_{\mu}({\cal{C}}_m) \}=0 \quad i,j=1,\dots,k+1, \quad l,m=1,\dots,r
$$
and by taking the residues of the functions $\Delta^{(i)}_{\lambda}({\cal{C}}_l)$, complete integrability is recovered. If the $\ep$ parameters are all distinct, then the residues of the function $\Delta^{(i)}_{\lambda}({\cal{C}}_l),\ i=1,\dots,N$ give simply the Casimir functions of the $i-$th copy of $A$.   

Alternatively, we can complete the set of integrals in involution through a different set of maps:
\begin{eqnarray}
&& \Delta^{(i-1)}_\la(y^\al)=\sum_{j=1}^{i-1} \frac{y^\al_j}{\la}+\frac{y^\al_{i}}{\la-\eta_1}, \qquad i=2,\dots,d_1 \nonumber\\
&&  \Delta^{(i-2)}_\la(y^\al)=\sum_{j=d_1+1}^{i-1} \frac{y^\al_j}{\la}+\frac{y^\al_{i}}{\la-\eta_1} \qquad i=d_1+2,\dots,d_2 \nonumber\\
&& \vdots \label{loopcop1} \\
&& \Delta^{(i-k)}_\la(y^\al)=\sum_{j=N-d_k+1}^{i-1} \frac{y^\al_j}{\la}+\frac{y^\al_i}{\la-\eta_1} \qquad i=N-d_k+2,\dots,N \nonumber\\
&& \Delta^{(N-k+1)}_\la(y^\al) \equiv \Delta_\la (y^\al)=\sum_{j=1}^{d_1} \frac{y^\al_j}{\la-\delta_1}+ \sum_{j=d_1+1}^{d_2} \frac{y^\al_j}{\la-\delta_2}+
\dots + \sum_{j=N-d_k+1}^{N} \frac{y^\al_j}{\la-\delta_k} +c^\al \nonumber
\end{eqnarray}
Such maps satisfy:
\begin{eqnarray*}
&& \pois{\Delta^{(i)}_\la(y^\al)}{\Delta^{(j)}_\mu(y^\be)}=0 \\
&& \quad {\rm if} \  d_k <i-k-2<d_{k+1}, \ d_l <j-l-2<d_{l+1} \quad k \neq l \\
&& \pois{\Delta^{(i)}_\la(y^\al)}{\Delta^{(j)}_\mu(y^\be)}=\frac{1}{\mu} C^{\al \be}_\ga  \, \Delta^{(i)}_{\lambda}(y^\ga)\\
&& \quad
{\rm if}\ d_k <i-k-2<d_{k+1}, \ d_k<j-k-2<d_{k+1} \quad j>i \\
&& \pois{\Delta^{(i)}_\la(y^\al)}{\Delta^{(i)}_\mu(y^\be)}=\frac{1}{\mu-\la} C^{\al \be}_\ga  \left( \Delta^{(i)}_{\lambda}(y^\ga)-
 \Delta^{(i)}_{\lambda}(y^\ga) \right) \quad i=1,\dots,N-k \\  
&& \pois{\Delta^{(i)}_\la(y^\al)}{\Delta^{(N-k+1)}_\mu(y^\be)}= \frac{1}{\mu-\delta_{i}} C^{\al \be}_\ga  \, \Delta^{(i)}_{\lambda}(y^\ga) \qquad i=1,\dots,N-k\\
&& \{ \Delta^{(N-k+1)}_{\lambda}(y^\al), \Delta^{(N-k+1)}_{\mu}(y^\be) \} = \frac{1}{\mu-\la} C^{\al \be}_\ga  \left( \Delta^{(N-k+1)}_{\lambda}(y^\ga)- \Delta^{(N-k+1)}_{\mu}(y^\ga)\right), 
\end{eqnarray*}
and they are also of the form required in Theorem 1. It follows that the images of the Casimir functions under this alternative set of maps give
another family of involutive functions that, to the best of our knowledge, has not been considered in the literature. Since this two families have in common the Gaudin Hamiltonians
$$
\Delta_\la({\cal{C}}_i), \qquad i=1,\dots,r
$$ 
it follows that the rational Gaudin model with degeneracies is superintegrable (see also \cite{Harnad}). The degree of superintegrability increase by increasing
the degeneracy of the parameters $\ep_i, \ i=1,\dots,N$.

\section{Generalization of the coproduct method to higher rank Lie--Poisson algebras} \label{higher}

As we have remarked in section \ref{coproduct}, the coproduct method is unable to provide complete integrability when applied to a simple Lie--Poisson algebra with a rank higher than $1$. In this section we will show how, using the loop coproduct formulation, it is possible to build up a larger set of
involutive functions with respect to those obtained through the coproduct; in the next subsection we will show that, in the case of simple Lie algebras,
these families are large enough to define completely integrable systems.
  
Again, we consider the case when $A$ is a Lie--Poisson algebra ${\cal{F}}(\alg^*)$.
We recall that the primitive coproducts are given by (\ref{standard}):
$$
\Delta^{(i)}(y^\al)=\sum_{j=1}^i y^\al_i \qquad i=2,\dots,N
$$
Let us introduce the following loop coproducts
\begin{equation}
\Delta_\la^{(i)}(y^\al)=\frac{\Delta^{(i-1)}(y^\al)}{\la}+ \frac{y^\al_i}{\la-\epsilon_1} \qquad  i=2,\dots,N \label{loopcop}
\end{equation}
where $\epsilon$ is a constant parameter ($\ep \neq 0$). 
Notice that these maps coincide with the first $N-1$ ones in (\ref{loopcop1}) when $k=1$ (that is when the parameters $\ep_i$ are all of them equal).
The loop coproducts (\ref{loopcop}) satisfy the Poisson brackets:
\begin{eqnarray}
&&   \{ \Delta_{\la}^{(i)}(y^\al), \Delta_{\mu}^{(k)}(y^\be) \}= \pois{\frac{\Delta^{(i-1)}(y^\al)}{\la}+ \frac{y^\al_i}{\la-\epsilon}}{\frac{\Delta^{(k-1)}(y^\al)}{\mu}+ \frac{y^\al_k}{\mu-\epsilon}}= \nonumber\\
&&   =\frac{1}{\mu} C^{\al \be}_\ga \left( \frac{\Delta^{(i-1)}(y^\ga)}{\la}+ \frac{y^\ga_i}{\la-\epsilon} \right)=
\frac{1}{\mu} C^{\al \be}_\ga  \Delta_{\la}^{(i)}(y^\ga)   
\end{eqnarray} 
which are of the form (\ref{p1}) with $f^\beta_\gamma(i,k,\la,\mu,\vec{Y})= 1/\mu$.

On the other hand, we have:
\begin{eqnarray}
&&   \{ \Delta_{\la}^{(i)}(y^\al), \Delta_{\mu}^{(i)}(y^\be) \}= \pois{\frac{\Delta^{(i-1)}(y^\al)}{\la}+ \frac{y^\al_i}{\la-\epsilon}}{\frac{\Delta^{(i-1)}(y^\al)}{\mu}+ \frac{y^\al_i}{\mu-\epsilon}}= \nonumber\\
&&   =C^{\al \be}_\ga \left( \frac{\Delta^{(i-1)}(y^\ga)}{\la \mu} + \frac{y^\ga_i}{(\mu-\epsilon)(\la-\epsilon)} \right)= \nonumber\\
&&   = \frac{1}{\mu-\la} C^{\al \be}_\ga  \left( \frac{\Delta^{(i-1)}(y^\ga)}{\la}+ \frac{y^{\ga}}{\la-\epsilon} -
\frac{\Delta^{(i-1)}(y^\ga)}{\mu}+ \frac{y^{\ga}}{\mu-\epsilon}\right)= \nonumber \\
&&  = \frac{1}{\mu-\la} C^{\al \be}_\ga  \left( \Delta^{(i)}_\la (y^\ga) -
\Delta^{(i)}_\mu (y^\ga) \right)   \label{lgau}
\end{eqnarray} 
which is of the form (\ref{p2}) with $g^\be_\ga(i,\la,\mu,\vec{Y})=-h^\al_\ga(i,\la,\mu,\vec{Y})= 1/(\mu-\la)$. Notice that (\ref{lgau}) coincide with the rational Gaudin algebra (\ref{Gaudinalgebra}).
From Theorem \ref{main} it follows that:
$$
\{ \Delta^{(i)}_{\lambda}({\cal{C}}_j) (\vec{y}),  \Delta^{(k)}_{\mu}({\cal{C}}_l) (\vec{y}) \}=0  
$$
Moreover, from (\ref{lambda1}), it follows:
\begin{equation}
\{ \Delta^{(i)}_{\lambda}({\cal{C}}_j) (\vec{y}) ,  \Delta^{(k)}(y^\al) \}=0  \qquad k \geq i \quad \forall \al  \label{copgen}
\end{equation}
and, in particular
$$
\{ \Delta^{(i)}_{\lambda}({\cal{C}}_j) (\vec{y}) ,  \Delta^{(N)}(y^\al) \}=0  \qquad i=2,\dots,N \quad \forall \al.
$$ 
This result means that we can add to the involutive functions generated through the loop coproducts of the Casimirs $\Delta^{(i)}_{\lambda}({\cal{C}}_j)$ a 
maximal abelian subalgebra of $\Delta^{(N)}(A) \simeq A$ in the same way as in the standard coproduct case. 

Now let us show that the standard coproduct Hamiltonians are functionally dependent on the Hamiltonians obtained through the loop coproduct.
The Casimir functions ${\cal{C}}_j (\vec{y})$ can be always taken to be homogeneous functions:
\begin{equation}
{\cal{C}}_j(\la y^1, \dots, \la y^M)=f_j(\la) \ {\cal{C}}_j(y^1, \dots,  y^M) \label{homogeneous}
\end{equation}
If we choose $\epsilon=0$, then we have
$$
\frac{\Delta^{(i)}_\la({\cal{C}}_j(\vec{y}))}{f_j(\la)}=\frac{{\cal{C}}_j(\Delta^{(i)}_\la(\vec{y}))}{f_j(\la)}=\frac{{\cal{C}}_j\left(\frac{\Delta^{(i)}(\vec{y})}{\la}\right)}{f_j(\la)}= {\cal{C}}_j(\Delta^{(i)}(\vec{y}))=\Delta^{(i)}({\cal{C}}_j(\vec{y}))
$$
So that we get back the standard coproduct Hamiltonians. Moreover, since 
$$
\{ \Delta^{(i)}_{\lambda}({\cal{C}}_j) (\vec{y}), \Delta^{(i)}({\cal{C}}_j) (\vec{y})\}=0
$$
by virtue of (\ref{homogeneous}), it follows that $\Delta^{(i)}({\cal{C}}_j) (\vec{y})$ must be functionally dependent on the Hamiltonians 
obtained through the generating function $\Delta^{(i)}_{\lambda}({\cal{C}}_j) (\vec{y})$.

\subsection{Complete integrability in the case of simple Lie-Poisson algebras} \label{simpleal}

Let us show that through the loop coproducts (\ref{loopcop}) it is always possible to define a completely integrable system 
when $A$ is a Lie-Poisson algebra ${\cal{F}}(\alg^*)$ associated with a simple Lie algebra $\alg$.
 
We recall from Section \ref{coproduct} that the number of degrees of freedom of an Hamiltonian system defined on 
$\overbrace{A \otimes \dots \otimes A}^N$ is 
\begin{equation}
d_N=\frac{N(M-r)}2
\end{equation}

Let us denote with ${\cal{C}}_j$ the fundamental Casimirs of $A$, 
which are homogeneous polynomials in the generators of degree $m_j+1$, where $m_j,\ j=1,\dots,r$ are the exponents of the Lie algebra $\alg$
\cite{Fuchs}.
Let us multiply $\Delta^{(i)}_{\lambda}({\cal{C}}^{(j)})$ by $\lambda^{m_j+1}$, then
\begin{eqnarray*}
&& \la^{m_j+1} \Delta^{(i)}_{\lambda}({\cal{C}}^{(j)})={\cal{C}}^{(j)}\left( \Delta^{(i-1)}(y^1)+\frac{\la}{\la-\epsilon}y^1_i, \dots, 
\Delta^{(i-1)}(y^M)+\frac{\la}{\la-\epsilon}y^M_i \right)= \\
&& =\Delta^{(i)}_{\tilde{\lambda}}({\cal{C}}^{(j)})(\vec{y})
\end{eqnarray*}
where 
\begin{equation}
\Delta^{(i)}_{\tilde{\lambda}}(y^\al)=\Delta^{(i-1)}(y^\al)+\tilde{\la} y^\al_i \qquad \tilde{\la}=\frac{\la}{\la-\epsilon} \label{base2}
\end{equation}
Let us expand $\Delta^{(i)}_{\tilde{\lambda}}({\cal{C}}^{(j)})$ in terms of $\tilde{\lambda}$
\begin{equation}
\Delta^{(i)}_{\tilde{\lambda}}({\cal{C}}^{(j)})= K_0^{(i,j)} + \tilde{\lambda} K_1^{(i,j)}+ 
\dots + \tilde{\lambda}^{m_j+1} K_{m_j+1}^{(i,j)} \label{svil}
\end{equation}
We notice that:
$$
K^{(i,j)}_0= \Delta^{(i-1)}({\cal C}^{(j)})= \sum_{l=0}^{m_j+1} K^{(i-1,j)}_l \qquad
K^{(i,j)}_{m_j+1}= {\cal{C}}^{(j)}_i
$$
so that in the set of functions
$$
\{K_{l}^{(i,j)}\} \qquad i=2,\dots,N \quad j=1,\dots,r \quad l=0,\dots,m_j+2
$$ 
at most $(N-1) \, \sum_{j=1}^r m_j$ are functionally independent among them and with respect to the
Casimirs. From
equation (\ref{lambda1}) it follows that
we can add a maximal abelian subalgebra of $\Delta^{(N)}(A) \simeq A$; that is at most $(l-r)/2$ further
involutive functions.

In conclusion, using the loop coproduct, we can define at most
$$
(N-1) \sum_{j=1}^r m_j+ (l-r)/2 
$$
functions in involution.
This set of functions will define an integrable system iff
\begin{equation}
2 \sum_{j=1}^r m_j \geq (l-r). \label{nec}
\end{equation}

 In the case of simple Lie algebras, formula (\ref{nec}) means that twice the sum of all exponents has to be greater or equal to the dimension minus the rank. Let us denote with $h$ the coxeter number associated 
with the simple Lie algebra, then the following two formulas holds:
\begin{eqnarray}
&& rh=l-r  \label{rh}\\
&& m_i+m_{r-i+1}=h \qquad \qquad \forall \, i \, \in \, (1,\dots,r)  \label{exponents}
\end{eqnarray}
Summing (\ref{exponents}) on all possible values of $i$ and using equation (\ref{rh}), we get:
$$
\sum_{i=1}^r m_i+m_{r-i+1}= 2 \sum_{i=1}^r m_i = r h= l-r
$$ 
So that, for simple Lie algebras, the necessary condition for integrability (\ref{nec}) is always verified. Moreover, it can be proven that the set of
functions $\{K_{l}^{(i,j)}\}$  provide indeed $ (N-1) \times \sum_{j=1}^r m_j$ functionally independent involutive functions (see for example \cite{RSTS}), since the counting of independent functions boils down to the counting of independent spectral invariants of the Lax matrix of the two-body homogeneous rational $\alg$-Gaudin model.
In conclusion, in the case of simple Lie algebras, if we can define an integrable system on the corresponding
Lie--Poisson algebra $A$, then, using the loop coproduct, we can define an integrable system on
$A^{\otimes N}$.

Finally, let us stress that an integrable system on a simple Lie--Poisson algebra can be always constructed through the spectral invariants
of the following Lax matrix:
$$
L(\la) \doteq \sum_{\al} X^\al y^\al + \la \sigma,
$$
where $\sigma \in \rho(\alg)$ must have simple spectrum. The $N$-th coproduct of these Hamiltonians coincide with the residues of 
$\Delta_\la^{(N+1)}({\cal C}_j), \ j=1,\dots,r$, where $\Delta_\la^{(N+1)}$ is the map in the last line of equation (\ref{loopcop1})
when $k=1$. 

\subsection{An example on a non-semisimple algebra} \label{twophoton}

We consider the two-photon Lie--Poisson coalgebra $(h_6, \Delta)$, that is a non-semisimple, rank two, six dimensional Lie--Poisson algebra
spanned by the six generators $\{N,A^+,A^-,B^+,B^-,M \}$ with Poisson brackets:
$$
\begin{array}{lll}
\{N,A^+\}=A^+, &\{N,A^-\}=-A^-, & \{A^-,A^+\}=M,\\
\{N,B^+\}=2 B^+, & \{N,B^-\}=-2B^- ,& \{B^-,B^+\}=2(M+2N),\\
\{A^+,B^-\}=-2 A^- ,& \{A^+, B^+\}=0, & \{M,\cdot\}=0, \\
\{A^-,B^+\}=2A^+ ,& \{A^-,B^-\}=0, &
\end{array}
$$    
$h_6$ has two fundamental Casimir functions: the mass $M$ and a third order Casimir given by
$$
{\cal C}= M(N^2- B^+ B^ - + MN)-A^+ A^-(M+2N)+B^+ (A^-)^2+ (A^+)^2 B^-. 
$$
The first step is to define an integrable system on $h_6$. 
Since $h_6$ is six dimensional with two Casimirs, we need two involutive functions.
A (rather trivial) possible choice for involutive functions on $h_6$ is: 
$$
f_1=B^+ + A^- + B^-, \qquad f_2=(A^+)^2+(A^-)^2+(A^- M).
$$
Notice that, from (\ref{base2}), (\ref{svil}), $\Delta_{\tilde{\lambda}}({\cal C})$ will be a third order polynomial in $\tilde{\lambda}$,
the coefficients being:
\begin{eqnarray*}
\tilde{\lambda}^0 : K^{(2)}_0 &=& {\cal C} \otimes 1\\
\tilde{\lambda}^1 : K^{(2)}_1 &=& M_1(2 N_1 N_2-B^{+}_1 B^{-}_2-B^{+}_2 B^{-}_1+M_1 N_2+M_2 N_1)+ \\
&+& M_2 (N_1^2+M_1 N_1 -B^{+}_1 B^{-}_1)- \\
&-& A^{+}_1 A^{-}_1 (M_2+2 N_2)-(A^{+}_1 A^{-}_2 + A^{+}_2 A^{-}_1)
(M_1+2 N_1)+\\
&+& 2 B^{+}_1 A^{-}_1 A^{-}_2+B^{+}_2 (A^{-}_1)^2+(A^{+}_1)^2 B^{-}_2+2 A^{+}_1 A^{+}_2 B^{-}_1\\
\tilde{\lambda}^2: K^{(2)}_2&=& 1 \leftrightarrow 2\\
\tilde{\lambda}^3: K^{(2)}_3&=&1 \otimes {\cal C}
\end{eqnarray*}
$K^{(2)}_2$ has the same expression as $K^{(2)}_1$ but with the indices $1$ and $2$ interchanged.
By construction $K^{(2)}_1$ and $K^{(2)}_2$ are in involution among them and with $\Delta(f_1)$ and $\Delta(f_2)$.
These four functions define an integrable system on $h_6 \otimes h_6$.

We can define the following Poisson (iso-)morphism between (the symplectic leaves of) $h_6 \otimes h_6$ and the $8-$dimensional symplectic algebra 
$\{q_i,p_j\}=\delta_{ij} \ i,j=1,\dots,4$:
\begin{eqnarray*}
D(A^+ \otimes 1)= \mu_1 p_1+ \mu_2 p_2  && D(1 \otimes A^+)= \mu_3 p_3+ \mu_4 p_4  \\
D(A^- \otimes 1)=\mu_1 q_1+ \mu_2 q_2 \ && D(1 \otimes A^-)=\mu_3 q_3+ \mu_4 q_4  \\
D(M \otimes 1)=\mu_1^2+\mu_2^2          &&  D(1 \otimes M)=\mu_3^2+\mu_4^2  \\
D(B^+ \otimes 1)=p_1^2+ p_2^2+ \frac{a_1}{(\mu_1 q_2 -\mu_2 q_1)^2} && D(1 \otimes B^+)=p_3^2+ p_4^2+ \frac{a_2}{(\mu_3 q_4 -\mu_4 q_3)^2}\\
D(B^- \otimes 1)= q_1^2+q_2^2 && D(1 \otimes B^-)=q_3^2+q_4^2 \\
D(N \otimes 1)=p_1 q_1+p_2 q_2 -\frac{\mu_1^2+\mu_2^2}{2} &&  D(1 \otimes N)=p_3 q_3+p_4 q_4 -\frac{\mu_3^2+\mu_4^2}{2}
\end{eqnarray*}
With this choice the functions
$$
\{ D(\co(f_1)), D(\co(f_2)), D(K^{(2)}_1), D(K^{(2)}_2) \}
$$
define an integrable system on the $8-$dimensional symplectic algebra. In particular, we have:
\begin{eqnarray*}
D(\co(f_1)) &=& \sum_{i=1}^4 \left( p_i^2 +\mu_i q_i + q_i^2 \right) + \frac{a_1}{(\mu_1 q_2 -\mu_2 q_1)^2} + \frac{a_2}{(\mu_3 q_4 -\mu_4 q_3)^2}\\
D(\co(f_2)) &=& \left( \sum_{i=1}^4 \mu_i p_i \right)^2+ \left( \sum_{i=1}^4 \mu_i q_i \right)^2+ \left( \sum_{i=1}^4 \mu_i^2 \right) 
\left( \sum_{i=1}^4 \mu_i q_i \right)
\end{eqnarray*}
The explicit expressions for $D(K^{(2)}_1)$ and $D(K^{(2)}_2)$ are really cumbersome, so that we do not write them\footnote{In \cite{alfonsoh6} a degenerate symplectic realization has been considered. In this case, to obtain an integrable system, one can indeed consider only the standard coproduct of the Casimirs but, at the same time, still has to begin with an integrable system defined on the abstract algebra $h_6$.}.

\section{Conclusions} \label{conclusion}

In this paper we have presented a generalization of both the Gaudin algebras and the coproduct method that we called loop-coproduct method.
We showed that this method provides a unified theoretical framework for rational, trigonometric and elliptic Gaudin models, and also for the integrable models coming from the coproduct method. Moreover we showed how, in the case of Lie-Poisson algebras, the loop-coproduct method makes it
possible to define larger sets of involutive functions with respect to the coproduct method and that this larger sets are indeed complete when 
the Lie-Poisson algebra is simple. In this section we would like to mention some open questions that, in our opinion, deserve further investigation.
One of the questions is whether this approach can be extended also to quantum systems, that is how one should reformulate theorem \ref{main} when 
$A$ is a noncommutative algebra (with respect to the multiplication). The quantization of the rational Gaudin Hamiltonians, found in \cite{Ta},
could be a starting point to investigate this problem.  

The only examples of integrable systems obtained from theorem \ref{main} that we discussed in this paper are related to Lie-Poisson algebras. On the other hand the theorem is much more general, since can be applied to arbitrary Poisson algebras. 
Indeed, as the coproduct method is a particular case of theorem \ref{main}, the integrable systems coming from the coproduct method applied to q-Poisson algebras (see \cite{Balrag}) also fall into this framework. However, these are the only examples that we know of integrable systems that can be obtained by applying the loop-coproduct method to a (non-linear) Poisson algebra and in all these examples there is no dependence on the ``spectral'' parameters $\la$ and $\mu$. Equation (\ref{p2}) can be seen as a defining equation for $\Delta^{(k)}_\la$. Indeed, it can be rewritten 
in the form
\begin{eqnarray}
&& \sum_i  \frac{ \partial \Delta_{\la}^{(k)}(y^\al)}{\partial y^\delta_i} \frac{\partial \Delta_{\mu}^{(k)}(y^\be)}{\partial y^\gamma_i} F^{\ga \de}(\vec{y_i})= \nonumber \\
&&= g^\beta_\ga(k,\la,\mu,\vec{Y}) F^{\al \ga} (\Delta_{\la}^{(k)}(\vec{y}))+ h^\al_\ga(k,\la,\mu,\vec{Y}) F^{\ga \be} (\Delta_{\mu}^{(k)}(\vec{y}))
\label{diffunc}
\end{eqnarray}
This is a non-linear PDE with a functional part in the spectral parameters $\la,\mu$. Clearly, finding a solution (if it exists) is a formidable
task. On the other hand, in equation (\ref{diffunc}), we have the freedom of choosing the Poisson algebra (that is the functions $F^{\al \be})$ and
the arbitrary functions $g^\beta_\ga$ and $h^\al_\ga$. A clever choice of such functions could make the solution of equation (\ref{diffunc}) 
affordable. In our opinion, it would be very interesting to find other maps defined on a Poisson algebra, depending on the spectral parameters and
satisfying the hypotheses of the theorem \ref{main}. Indeed, such maps would define a natural generalisation of the Gaudin algebras from the Lie-Poisson to the Poisson algebra context and could, hopefully, provide examples of new integrable models. 

\section*{Acknowledgments}

The author is pleased to thank A. Ballesteros and O. Ragnisco for precious discussions and text revision. This work was partially supported by the Spanish MICINN under grant MTM2007-67389 (with EU-FEDER support), by Junta de Castilla y Le\'on (Project GR224) and by INFN-CICyT.

\end{document}